# Quantum-coherent light-electron interaction in an SEM

R. Shiloh†, T. Chlouba†, and P. Hommelhoff

Physics Department, Friedrich-Alexander-Universität Erlangen-Nürnberg (FAU), Staudtstraße 1, 91058 Erlangen, Germany

**The last two decades experimentally affirmed the quantum nature of free electron wavepackets by the rapid development of transmission electron microscopes[1] into ultrafast, quantum-coherent systems[2–9]. In particular, ultrafast electron pulses can be generated and timed to interact with optical near-fields, yielding coherent exchange of the quantized photon energy between the relativistic electron wavepacket and the light field. So far, all experiments have been restricted to the physically-confining bounds of transmission electron microscopes, with their small, millimeter-sized sample chambers. In this work, we show the quantum coherent coupling between electrons and light in a scanning electron microscope[10], at unprecedentedly low electron energies down to 10.4 keV, so with sub-relativistic electrons. Scanning electron microscopes not only afford the yet-unexplored electron energies from ~0.5 to 30 keV providing optimum light-coupling efficiencies[11–16], but they also offer spacious and easily-configurable experimental chambers for extended and cascaded optical set-ups, potentially boasting thousands of photon-electron interaction zones. Our results unleashes the full potential of quantum experiments including electron wavepacket shaping[17] and quantum computing with multiple arithmetic operations[18] and will allow imaging with low-energy electrons and attosecond time resolution.**

While transmission electron microscopes (TEMs) are built to provide highly coherent electron beams for diffraction-based imaging methods[19], they are large and costly both in terms of purchasing and maintenance. In contrast, scanning electron microscopes (SEMs) are built to image surfaces based on scanning a small, focused electron beam over a sample, not requiring electron beam coherence or large beam energies (70 – 300 keV in TEMs), making the electron column more compact, the device easy to operate and more affordable. The different *modi operandi* lead to entirely different sample chamber geometries and placements: in a typical SEM, the sample chamber measures 30 cm across, whereas it is 3-7 millimeters in a TEM.

A new mode of imaging and photon-electron interaction has been introduced more than a decade ago in TEMs, termed photon-induced near-field electron microscopy (PINEM)[3], which laid the foundation not only to a new ultrafast time-resolved electron imaging mode but also to breakthrough electron-light coupling results[5–7,20,21] and, more recently, to extremely high coupling strengths using cavities, whispering-gallery modes, quasi-phase matching and microresonators[9,22–25].

In PINEM physics, a single electron wavepacket and the light field exchange an integer number of photons in an inelastic interaction. In each such photon exchange event, the electron can absorb or emit one or more photons, which results in acceleration or deceleration of the electron. Multiple such events are possible because the electron energy is several orders of magnitude larger than the net change the photon energy induces. While the underlying theory is well-established[5,6,17,20], new predictions are still being made[17,26–29]. It revolves around the near-field interaction coupling constant, or PINEM field parameter g,

$$g(x,y) = \frac{e}{\hbar\omega} \int_{-\infty}^{\infty} E_z(x,y,z^{'}) e^{-iz^{'}\omega/v} dz^{'}, \qquad (1)$$

where e is the electric charge, $\hbar$ the reduced Planck constant, $\omega$ the central angular frequency of the laser field, $E_z$ the complex optical electric field projection in the direction of the electron propagation $\hat{z}$, and v the electron velocity. This expression is valid if the electron is measured after the interaction is finished, and is in fact the Fourier transform of the electric field's z-component with spatial frequency $\omega/v$. It implies that PINEM imaging with TEM energies will be more sensitive to low spatial-frequency near-fields decaying farther away from the sample boundary, however, with a relatively low photon-electron coupling. Conversely, in SEMs, slow electrons of around 10 keV generate shorter-length near-fields with higher coupling, a prerequisite to reach the optimum photon-electron coupling strength[11,14].

Imaging using photon-electron interactions includes experimentally investigating the quantum nature and the response of nano-structures, molecules and atoms to incident light, mediated by the generated evanescent electromagnetic fields[30–32]. Examples on the fundamentals of electron-light coupling include attosecond quantum coherent control[7,21,33], quantum state reconstruction[34] and generation[35], attosecond pulse generation[34,36,37], and photon statistics reconstruction[9].

These remarkable demonstrations are all based on (1) steering and positioning of high quality electron beams with nanometer resolution – the core features of electron microscopes – and (2) the ability to measure the electron's energy after its interaction with the optical near-field, with a resolution better than the photon energy of the driving light field. In TEMs, this is straightforward to do as electron energy loss spectrometry (EELS) is a widely-used measurement modality, so that the required spectral resolution can be achieved with standard, commercial EELS spectrometers. In contrast, this mode of imaging does not exist in normal SEM operation, which relies on signals detected from secondary- or back-scattered electrons, rather than transmitted electrons.

We equipped a commercial but significantly modified SEM with a specially designed and home-built compact high resolution magnetic spectrometer to introduce the required spectral energy resolution to the SEM (Figure 1). The design is based on an Omega energy filter[38], enabling us to observe the coherent energy transfer between the exciting 1030 nm laser light (photon energy of 1.2 eV) and our subrelativistic electrons. We focus the laser beam on a sharp tungsten needle tip to generate an optical near-field and measure the energy spectrum following the electron interaction with it.

Figure 1 shows a sketch of the experimental setup: an electron wavepacket (green), released from the SEM emitter following photoexcitation by an ultraviolet laser pulse (purple) and accelerated to 17.4 keV, travels in the $\hat{z}$ direction and interacts with the near-field of the tungsten needle tip (Figure 1inset) aligned to the y-axis. The near-field is generated at the tip by a 1030-nm, 250-fs pulsed laser beam (dark red) propagating along the x-axis. After the photon-electron interaction, the electron wavepacket is dispersed in our photon energy-resolving spectrometer, the dispersion plane of which is imaged on a micro-channel plate detector with phosphor screen, allowing single electron detection (see Methods for more details).

Figure 2 shows photon-resolved electron spectra. We focused the electron beam to various positions in close proximity of the needle tip, such that the beam samples the near-field, and recorded a series of spectra (black line with blue background) at different locations along the tip axis indicated by the colored circles (Figure 1a). Both the number and the amplitude of the photon orders vary as a function of electron beam position: if the beam electrons pass the tip outside of the near-field, no PINEM orders are observed (panel b). Steering the electron beam up the tip, we observe an increase in PINEM orders from 3 on each side (panel c) up to 7 (panel f). Simultaneously, the spectral amplitudes vary. Numerical simulation results (blue lines with red background) based on standard PINEM theory almost perfectly match these results: First, we calculate the near-field coupling parameter $g$ of the needle tip from a 3D electromagnetic field simulation (Figure 2a). From this, it is straightforward to obtain the PINEM orders (see Methods). We note that the simulation does not take into account other effects that might affect the spectra, in particular electron energy loss by Smith Purcell radiation, for example; indeed, a slightly enhanced signal towards energy loss is apparent in the measured data (Figure 1 and Figure 2).

Intriguingly, the number of PINEM orders observed here is similar to those in a TEM under similar conditions, i.e., with an individual nano-object generating the near-field[7]. The largest photon order number we observed was 12, corresponding to a modulated photon energy of 14.4 eV (inset of Figure 1). It is noteworthy that we can also measure at least 4 photon orders at an energy as low as 10.4 keV (inset of Figure 2a), which we expect is by no means the lower limit. Future work with improved electron current stability will allow longer measurement sessions and by consequence a better signal to noise ratio (see Methods for more details). By focusing the electron probe to smaller dimensions, the sampling of the near-fields and accordingly the photon-electron coupling can be improved, yielding additional photon orders. Using high efficiency phase-matched coupling structures might allow measurement of thousands of PINEM orders as already observed with a TEM today[39]. Most interestingly, specially-designed structures and laser pulses can be engineered to shape the PINEM spectra[17]. Because of our single-electron detection efficiency, we can observe the build-up of the energy spectra on the detector in real-time (see Extended Data Movie).

The volume of the main sample chamber of a standard SEM typically measures at least 20 x 20 x 20 cm$^3$, versus the 3 mm$^3$ of a TEM. This offers a large flexibility in modifying the SEM chamber by installing large optical windows or electrical connections, and simplifying the coupling and extraction of signals from the vacuum chamber, a venture that can hardly be undertaken in a TEM. Furthermore, and even more important, an SEM can potentially accommodate many interaction locations for complex, cascaded quantum experimental schemes. A great practical advantage of an SEM is the convenience of using high numerical aperture optics (NA ≈ 1), which is challenging in a TEM. Very recently, nanophotonic chips have been introduced to couple light and electrons efficiently[9,25]. With even longer versions of these, we foresee up to tens of thousands independent light-electron interaction zones, which might enable complex free electron-based quantum computing[18]. Today, quantum information systems use the interplay of light and matter fundamentally tied to the energy levels of bound electrons, so in bulk materials or gases. In contrast, systems based on free electrons provide a quantized energy ladder accessible in high-energy and ultrashort time scales. In Ref. [18], the authors discuss the feasibility and provide a framework for such a quantum information manipulation system, and explain how the PINEM interaction can be employed to demonstrate a full single qubit. A second, most important building block in that scheme is free-space propagation of the free electrons. Several such blocks are required to implement qubit algebra, an experimental feat that may enjoy the spacious benefits of a photon-energy resolving SEM. With these abilities, such an SEM is perfectly staged for quantum-coherent experiments, including arbitrary electron wavefunction shaping[29], and our demonstration of the feasibility of utilizing an SEM to perform such endeavors is a first, invaluable step in this direction. Having access to the distribution of discrete energy levels of an electron wavepacket, both in terms of manipulation and observation, is paramount to fundamental research in the quantum nature of matter, but also as a tool in future applications of quantum information and theory, and as a tool for ultrafast time-resolved electron imaging applications. Finally, the ideal energy for various exciting future quantum-coherent electron-matter coupling experiments such as free electron bound electron interaction (FEBERI)[13] is in the range of a few keV[15,16], easily attainable with an SEM.

**Figures**

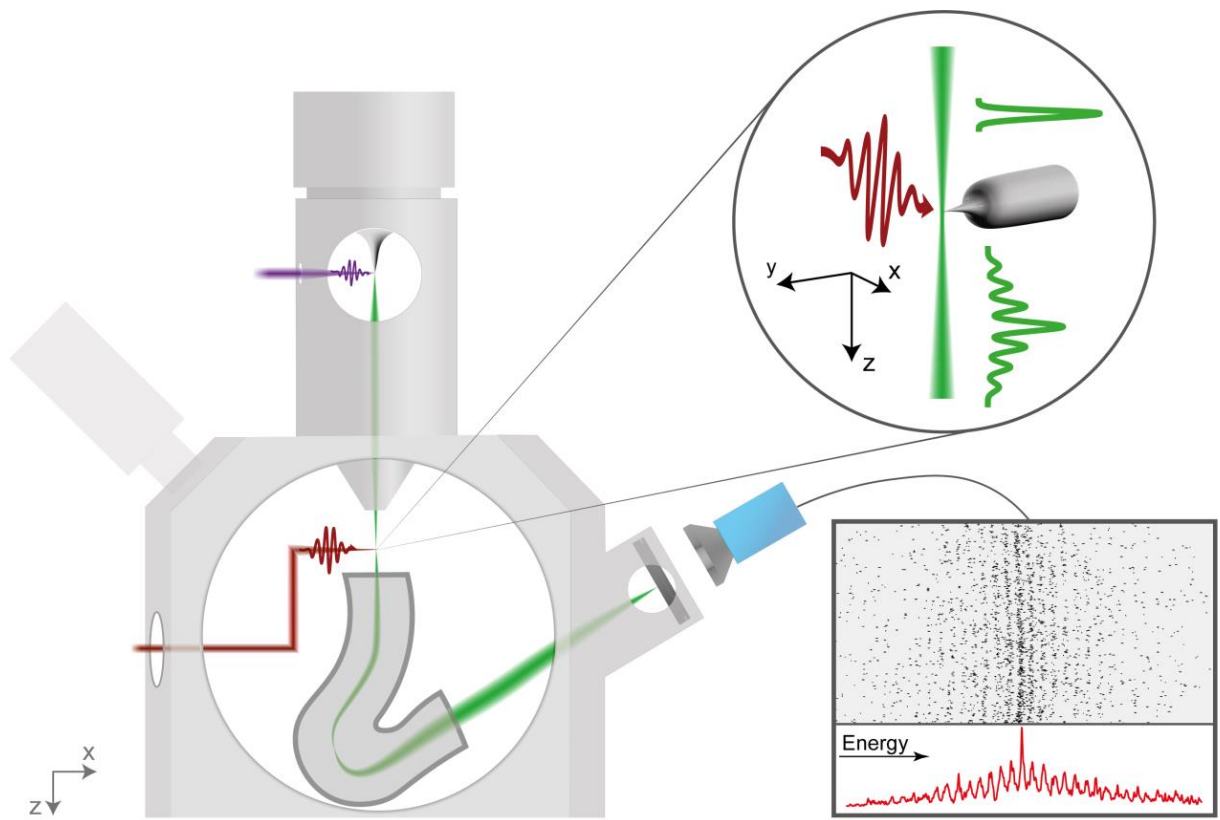

Figure 1 | **Quantum coherent electron-light coupling in an ultrafast SEM.** Electrons photoemitted by ultraviolet laser pulses (purple) propagate through the standard column of a commercial SEM. The electron beam (green) is focused close to a tungsten needle tip (inset), where it interacts with the optical near-field of the needle tip, excited by 1030-nm laser pulses, coupled into the SEM through a CF-100 window in the SEM sample chamber. The aspherical focusing lens (not shown) is 25 mm away from the tip, inside of the sample chamber. Electron spectra are recorded with a home-built compact double-stage magnetic sector electron spectrometer based on the Omega filter, placed inside of the SEM. The dispersion plane of the spectrometer is imaged onto a microchannel plate detector, whose phosphor screen is optically recorded from outside of the vacuum chamber with a CMOS camera. An example image is shown in the bottom right inset, where individual electron counts (black dots) and photon orders (vertical dotted lines) can be easily seen by eye. The PINEM spectrum is obtained by integrating the camera image vertically (see Methods). The experimental spectrum (red) shows 24 PINEM orders, 12 on each side, the maximum we observed.

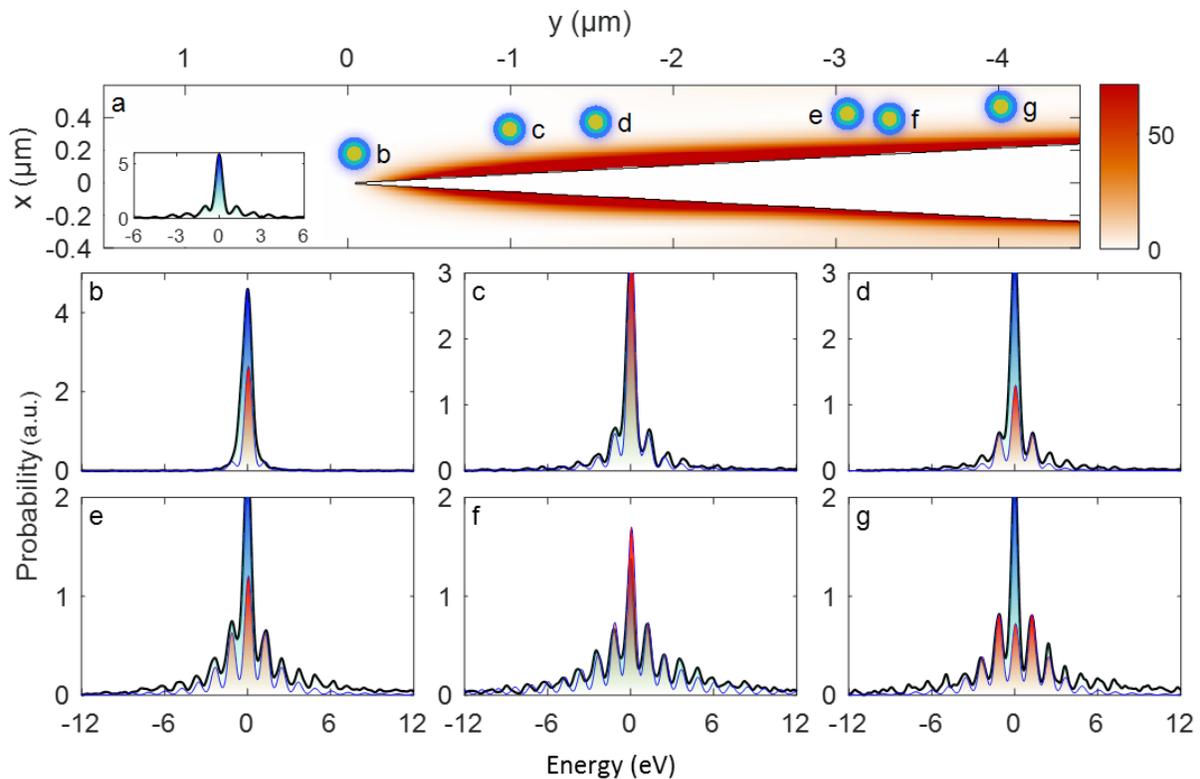

Figure 2 | **PINEM spectra as function of position along the needle tip.** (b - g) Six electron energy spectra at a central beam energy of 17.4 keV, recorded and matched to positions indicated by the colored circles in (a). The color scale in (a) shows a simulation of the near-field coupling parameter *g*. Spectrum (b) does not show photon orders, as expected for a beam not passing the optical near-field of the tip. Spectrum (c) shows 3 photon orders on either side of the zero loss peak, spectrum (d) 4, and up to 7 photon orders in spectrum (f). Experimental data are

displayed by the thick black curve (blue background), while the numerical results are shown with blue curves (red background). Clearly, the numerical results match the experimental data very well. The height of the zero-loss peak depends on the laser and electron pulse lengths, which was ignored in the simulation; instead, the amplitude of the first positive simulated energy peak was normalized to the experimental one. The positions of the circles in (a) are found such that the incoherent average of the simulated PINEM spectra match the experimental ones best. For details, see text and Methods. The inset in (a) shows PINEM spectrum recorded with 10.4 keV electrons. Also at this low energy we observe 4 photon orders on each side.

## Funding

We gratefully acknowledge funding by the Gordon and Betty Moore Foundation (#GBMF4744) and ERC Grant AccelOnChip (#884217).

## Competing interests

Authors declare no competing interests.

## Author contributions

T.C. measured the data, R.S. designed and built the electron spectrometer, and performed the simulations. R.S. and T.C. analysed the data. All authors contributed to the writing of the manuscript. P.H. supervised the experiment.

## Additional information

† R.S. and T.C. contributed equally to this work.

Corresponding authors: Roy Shiloh (roy.shiloh@fau.de), Tomáš Chlouba (tomas.chlouba@fau.de) and Peter Hommelhoff (peter.hommelhoff@fau.de).